\documentclass[conference]{IEEEtran}

\makeatletter
\def\ps@headings{%
\def\@oddhead{\mbox{}\scriptsize\rightmark \hfil \thepage}%
\def\@evenhead{\scriptsize\thepage \hfil \leftmark\mbox{}}%
\def\@oddfoot{}%
\def\@evenfoot{}}
\makeatother

\usepackage{cite}

\ifCLASSINFOpdf 

\else

\fi

\usepackage{paralist,color}
\usepackage{graphicx}
\usepackage{algorithm2e}
\usepackage{algorithmic}
\newtheorem{theorem}{Theorem}
\newtheorem{lemma}[theorem]{Lemma}

\newtheorem{definition}[theorem]{Definition}

\newcommand{\nix}[1]{}

\begin{document}
%
\title{Secure Hop-by-Hop Aggregation of End-to-End Concealed Data in Wireless Sensor Networks}

\author{\authorblockN{Esam Mlaih}
\authorblockA{Department  of Computer Science \\
Texas A\&M University, TX 77843, USA \\
Email: mlaih@tamu.edu} \and \authorblockN{Salah A. Aly}
\authorblockA{Department  of Computer Science \\
Texas A\&M University, TX 77843, USA \\
Email: salah@cs.tamu.edu}}

\maketitle

\begin{abstract}
In-network data aggregation is an essential technique in mission
critical wireless sensor networks (WSNs) for achieving effective
transmission and hence better power conservation. Common security
protocols for aggregated WSNs are either hop-by-hop or end-to-end,
each of which has its own encryption schemes considering different
security primitives. End-to-end encrypted data aggregation protocols
introduce maximum data secrecy with in-efficient data aggregation
and more vulnerability to active attacks, while hop-by-hop data
aggregation protocols introduce maximum data integrity with
efficient data aggregation and more vulnerability to passive
attacks.

In this paper, we propose a secure aggregation protocol for
aggregated WSNs deployed in hostile environments in which dual
attack modes are present. Our proposed protocol is a blend of
flexible data aggregation as in hop-by-hop protocols and optimal
data confidentiality as in end-to-end protocols. Our protocol
introduces an efficient $O(1)$ heuristic for checking data integrity
along with cost-effective heuristic-based divide and conquer
attestation process which is $O(\ln{n})$ in average -$O(n)$ in the
worst scenario- for further verification of aggregated results.

\end{abstract}

\section{Introduction}
A wireless sensor network is usually a collection of hundreds or
thousands of resource-constrained devices with small memories, low
bandwidth and limited power resources. They are deployed in fields
where persistent human monitoring and surveillance are either
impossible or infeasible. These small detectors can be used to sense
events ranging from simple readings (e.g. sensing room temperature)
to more important and sensitive measures (e.g. intruder detection in
military applications, detecting wildfire or signs of any
catastrophic phenomena). Raw data collected using these limited
sensors are usually queried by a more powerful device called base
station (BS) -which may be far away from sensing fields- for further
analysis and event-based reactions~\cite{stojmenovic05}.

Since wireless sensor networks are energy constrained and bandwidth
limited, reducing communications between sensors and base stations
has a significant effect on power conservation and bandwidth
utilization~\cite{Intanagonwiwat02}.  Aggregated sensor networks
serve this purpose by introducing designated nodes called
aggregators that provide efficient data collection and transmission.
An aggregator can sense its own data while aggregating received
results from children nodes, which in turn may be leaf sensors or
aggregators as well.

Aggregated wireless sensor networks provide better power
conservation and efficient use of communication channels but
introduce additional security concerns. A passive adversary may
capture sensitive results of aggregated data that represents a large
partition of the aggregated WSN if the key of the root aggregator of
that partition is compromised. On the other hand, an active
adversary can forge aggregated data of a partition by compromising
the parent node of that partition. Many security protocols for
aggregated WSNs were introduced to solve these security problems.
These security protocols can be classified according to their
underlying encryption schemes into end-to-end and hop-by-hop secure
data aggregation protocols.

The paper is organized as follows. In
Section~\ref{sec:related_work}, we present previous work on secure
aggregation on WSNs and we define our problem. In
Section~\ref{sec:model}, we present our network model and its design
goals, along with attacker model. In Sections~\ref{sec:protocol}
and~\ref{sec:sum}, we demonstrate our security protocol and provide
analysis of its complexity. The paper is concluded in
Section~\ref{sec:conclusion}.

\section{Related Work}\label{sec:related_work}
In this section, we give a short background on previous work of
secure aggregation protocols in WSNs,  which are classified as
end-to-end and hop-by-hop.

In end-to-end encryption
schemes~\cite{Castelluccia05,girao05,mykletun06,Sang06},
intermediate aggregators apply some aggregation functions on
encrypted data which they can't decrypt. This is because these
intermediate aggregators don't have access to the keys that are only
shared between data originators (usually leaf sensor nodes) and the BS.
In CDA~\cite{girao05} sensor nodes share a common symmetric key with the
BS that is kept hidden from middle-way aggregators. In~\cite{Castelluccia05} each
leaf sensor share a distinct long-term key with the
BS. This key is originally derived from the master
secret only known to the BS. These protocols show that aggregation of
end-to-end encrypted data is possible through using additive Privacy Homomorphism (PH)
as the underlying encryption scheme. Although these protocols are
supposed to provide maximum data secrecy across the paths between
leaf sensor nodes and their sink, overall secrecy resilience of a WSN
becomes in danger if an adversary gains access to the master key in~\cite{Castelluccia05},
or compromises only a single leaf sensor node in CDA to acquire the common symmetric key shared between all
leaf nodes.

In~\cite{mykletun06,Sang06} public key encryption based on elliptic
curves is used to conceal transient data from leaf sensors to the
BS. These schemes enhance secrecy resilience of WSNs against
individual sensor attacks, since compromising a single or a set of
sensor nodes won't reveal the decryption key that only the BS knows.
An attracting feature of \cite{mykletun06} is the introduction of
data integrity in end-to-end encrypted WSNs through Merkle hash
trees of Message Authentication Codes (MACs). However, both schemes
raise power consumption concerns, since computation requirements for
public key encryption is still considered high for WSNs
\cite{perrig01}.

Many hop-by-hop aggregation protocols in WSNs like
~\cite{chan06,hu03,Mahimkar2004,przydatek03,yang06}, provide more
efficient aggregation operations and highly consider data integrity.
However, since sensed data being passed to non-leaf aggregators are
revealed for the sake of middle-way aggregation, hop-by-hop
aggregation protocols represent weaker model of data confidentiality
perspective than end-to-end aggregation protocols. Data secrecy can
be revoked of a partition if a passive adversary has obtained the
key of the root aggregator of that partition.

\subsection{Problem Statement}
The challenge is to find a general security protocol for aggregated
WSNs that is not limited to certain topology and provides strong
data confidentiality comparable to those in secure end-to-end
communication protocols. Also, it can provide efficient data
aggregation and integrity comparable to those in hop-by-hop
aggregation, taking into account the presence of active and passive
adversaries. So, when some nodes of the aggregated WSN are
physically compromised, compromiser must not gain more information
or have influence on aggregated results beyond the effects of its
compromised nodes. For these purposes, we propose our security
protocol that provides end-to-end data concealment using data
diffusion, and in the same time, it provides secure and flexible
hop-by-hop aggregation with efficient data integrity test followed
by attestation process when forged data are detected in order to
eliminate and exclude contributions of any compromised nodes that
might be the source of the forged data.

\section{System Model}\label{sec:model}

\subsection{Notations}

We use the following notations to describe our protocol:

\begin{compactitem}
\item BS refers to the Base Station.
\item
$ S = \{S_1,S_2,\dots,S_n\} $ represents the set of
sensor/aggregator nodes in the WSN. Since in our model sensors have
the aggregation capabilities, the term sensor will be used to refer
to a sensor that aggregates as well.
\item $ID_{S_i}$ refers to the node ID of sensor node $S_i$.
\item $K_{S_i,S_j}$ denotes a pairwise symmetric key between node $S_i$ and node $S_j$. $K_{S_i}$ and $K^{'}_{S_i}$
 are two pairwise symmetric keys of node $S_i$ shared with the BS, and $\mathcal{K}$ is a set of all keys.
\item $m_{S_i}$ denotes a sensed data read by sensor $S_i$. $m_{S_i}$ is a bounded real value, i.e. $m_{S_i} \in \mathcal{D}=[u,v]$ for  maximum and minimum  sensible values $v$ and  $u$, respectively.
\item $Enc_K(m)$ denotes an encryption of a message $m$ using a key $K$.
\item $MAC(K_{S_i},m_{S_i})$ denotes a message authentication code of $m_{S_i}$ that is sensed by sensor $S_i$, this code is generated using the symmetric key $K_{S_i}$ that is shared between $S_i$ and the BS.
\item $F_K(m)$ refers to a diffusion algorithm that is a public knowledge in the WSN. It takes as input a key $K$ and a data $m$, the result is a diffused value $D \in [u,v]$.
\item $S_i \longrightarrow S_j$ represents a one (or more) hop communication from sensor node $S_i$ to $S_j$.
\end{compactitem}

\begin{figure}[t]
\centering
\includegraphics[scale=0.85]{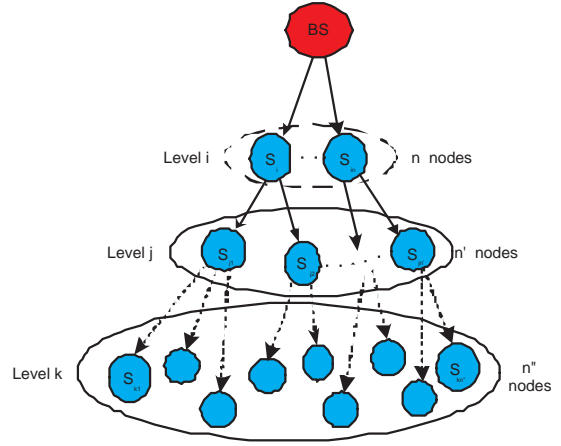}
\caption{Network Model (Aggregated WSN).}
\label{fig:netmodel}
\end{figure}
\subsection{Network Model}
We assume a general aggregated multi-hop WSN consisting of a large
collection of resource-constrained sensor/aggregator nodes (MICA
motes~\cite{mica2} for example) connected in a tree topology rooted
at a powerful node called the Base Station (BS). An illustration of
this model is depicted in Fig.~\ref{fig:netmodel}.
 We don't impose any restrictions on the topology as long as it
is a connected tree rooted at the BS. We don't require a specific
aggregation tree construction algorithm, any efficient tree
construction algorithm like TaG~\cite{madden02} can be used in our
model. The BS may initially issue aggregation queries or it may be
connected to an off-network distant querier which is in this case
considered data consumer, and the BS is considered its query server.
Aggregation queries represents the union of all sensor readings
along the paths of the WSN to its root, i.e. the BS.

We assume that every sensor node $S_i$ is deployed with two unique
symmetric keys $K_{S_i}$ and $K^{'}_{S_i}$ shared with the BS, using
a secure key deployment protocol, like MIB~\cite{PerrigLukKuo2007}.
A secure broadcast authentication protocol is assumed for
authenticating messages, an example of such protocol is
$\mu$TESLA~\cite{perrig01}. Secure key distribution between adjacent
nodes is also assumed, some can be found in ~\cite{Chan05}.

\subsection{Attacker Model}

We assume a dual operational mode adversary (both passive and
active) who is interested in revealing in-network data secrecy and
injecting forged data. In our model, we consider effective attacks,
where an adversary physically compromises $k \ll n$ nodes to gain
the advantage that would result of attacking $m$ nodes where $k < m
\leq n$ without the need of attempting such attack on these $m$
nodes directly. That is, with few compromised nodes, an adversary
can endanger the security of an aggregated WSN as if it had
physically compromised much larger collection of nodes. When we
denote a node as being physically compromised, we mean that an
adversary gained control over the node's operation, having access to
all its memory, keys, and resources, and is capable to reprogram
such a compromised node with attacking code. Attacker is not limited
to a single place, it can compromise scattered partitions of nodes
in which every partition may have nodes in parent/children
relationship.

In this work, we don't consider preventing attacks that disrupt the
regular operation of a WSN such as denial-of-service (DoS)
attack~\cite{Raymon08} or underlying routing protocol attacks. We
are interested in preventing attacks that aim to acquire aggregation
results or tamper them rather than attacks that aim to prevent a
querier from being served.

\subsection{Design Goals}

We designed our protocol to protect against spy-out and false data
injection attacks, for that, we considered the following security
perspectives:

\begin{itemize}
\item
Resilience: An adversary who compromises few nodes of an aggregated
WSN must not spy-out or gain any impact on the final aggregation
outcome beyond the influence of the readings and results of its
compromised nodes.

\item
Efficient Data Integrity, Commitment and Attestation: Aggregation
result must be verified to be the authentic union of sensor readings
and intermediate results. Such verification and attestation
processes should not impose significant overhead over the WSN that
is over aggregation communication overhead.

\item
Generality: The protocol should apply to any aggregated WSN with
arbitrary tree topology, moreover, the protocol should support
expandable WSNs without any extra reconfiguration.

\item
Status Monitoring: BS must determine when a sensor node becomes dead
or unreachable, by knowing and maintaining a list of all nodes
contributed in every aggregation query.

\end{itemize}

\section{Efficient and Secure Data Aggregation
Protocol}\label{sec:protocol}
In this section, we present our proposed protocol that resolves the
compromise between data secrecy and efficient aggregation. An
overview of the protocol will be presented first, then it will be
followed by discussing the protocol details.

\subsection{Overview}
Our protocol is designed over the approach of data diffusion that
preserves the mathematical relationships between different values
which are all bounded by a defined range. By preserving mathematical
relationships we can perform efficient hop-by-hop aggregation of
collected diffused data. The information of these mathematical
relationships are kept concealed end-to-end to maintain complete communication path
secrecy. Beside maintaining the mathematical relationships, the
diffusion algorithm must not increase the size of encrypted data.
Based on this, we can achieve efficient secure hop-by-hop
aggregation of end-to-end concealed data in aggregated WSNs.

\subsection{Network Setup and Query Dissemination}

After field deployment, communication paths should be established.
An efficient algorithm like TaG~\cite{madden02} can be used for tree
topology construction.  Communication channels are secured using
pairwise encryption keys between every parent/children nodes, this
is the same technique used in many hop-by-hop protocols (e.g.
\cite{yang06}) for securing communication channels.

After tree construction, every sensor node $S_i$ sends its
$ID_{S_i}$ and an initial random reading $m^0_i \in [u,v]$ to the BS
in a message encrypted using pairwise symmetric key $K_{S_i}$. The
initial random reading $m^0_i$ serves in data diffusion algorithm as
we will see later.

When the BS receives a query from a querier, it disseminates this
query through the WSN paths. This query contains the desired aggregation
function to be performed.

\subsection{Data Diffusion}

The purpose of the data diffusion process is to consolidate
transient data from intermediate aggregators while giving them
flexibility and efficiency while applying aggregation functions on
these concealed data. Data diffusion serves also in data integrity
check as we will see later. Every sensor node diffuses its sensed
data before transmission. Middle-way aggregation of diffused data
occurs before the final result reaches the BS, which is the only one
who can revert diffused result to its actual value.

Assume $S=\{S_1,S_2,\ldots,S_n\}$ be the set of sensor nodes and
every node $S_i$ reads a value $m_{S_i}$. Every sensor node $S_i$
uses a diffusion function $F_{K}(m_{S_i})$, using the keys $K_{S_i}$
and $K'_{S_i}$ to generate a pair of diffused data, where
$K_{S_i},K_{S_i}^{'}$ are two shared keys between $S_i$ and the base
station (BS). We define the diffusion function
$F_{K_{S_i}}(m_{S_i})$ as follows:
\smallskip

\begin{definition}
Assume  $PS: \mathcal{D} \times \mathcal{K} \rightarrow \mathcal{D}$
be a public generator map (i.e., one way function) to produce
\begin{equation}
D_j=PS(K_{S_i},D_{j-1})
\end{equation}
where $D_j \in \mathcal{D}$, $D_0=m_{S_i}^0$, and $K_{S_i} \in
\mathcal{K}$ for $j \geq 1$ and $1 \leq i \leq n$. Let $F:
\mathcal{D}\times \mathcal{D} \longrightarrow \mathcal{D}$ be a
diffusion function defined as
\begin{equation}
F_{K_{S_i}}(m_{S_i}^j) = PS(K_{S_i},D_{j-1}) \odot m_{S_i}^j,
\end{equation}
\end{definition}
The value of the generator sequence $PS$ is taken as an input along
with the sensed reading $m_{S_i}^j$ to the mathematical operand
$\odot$ which generates a diffused value $F_{K_{S_i}}(m_{S_i}^j) \in
\mathcal{D}$. There is no strict definition of operand $\odot$, it
refers to any reversible operation that takes two inputs and
produces an output that belongs to $\mathcal{D}$.
 Examples of $\odot$ could vary between trivial operators such as simple
addition "+" to more complex bijection functions. $D_j^{'}$ is
generated symmetrically to $D_j$, but using key $K^{'}_{S_i}$
instead of $K_{S_i}$.

Since the BS shares the private key $K_{S_i}$ and initial random
reading $m^0_{S_i}$ of every sensor node $S_i$,  the BS is able to
generate the diffusion value $D^j$ of every transmission phase. This
means that the BS can revert every diffused reading $D_{S_i}^j$ sent
by a sensor $S_i$ in the WSN to its actual value.

\begin{figure}[t]
\centering
\includegraphics[scale=0.8]{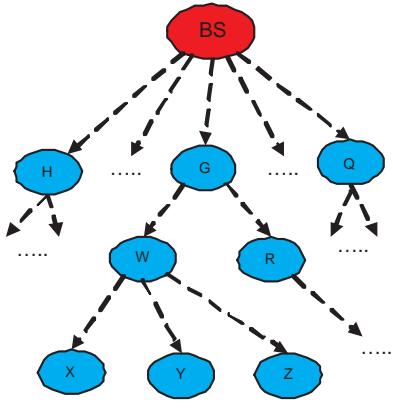}
\caption{An example of an aggregated WSN tree.}
\label{fig:sample_net}
\end{figure}
\section{The $SUM$ Aggregation}\label{sec:sum}
In this section, we propose the $SUM$ aggregation function in our
secure aggregation protocol. The algorithm that performs the $SUM$
aggregation SumAgg is illustrated in algorithm~\ref{alg:SumAgg}.
When the BS receives a query of $SUM$ aggregation function, it
broadcasts this request through the WSN. Whenever a sensor node gets
this request, it passes such a request to its children nodes, this
goes on until reaching leaf level. A leaf sensor node receiving this
request will send its diffused reading to its parent. For
illustration purposes, let us consider the network in
Fig.~\ref{fig:sample_net}. Leaf sensor $X$ sends the following
packet to its immediate parent $W$:
\begin{eqnarray}\label{eq:fromWtoX}
X  \longrightarrow W&:& ID_X,IV_{X,W},Enc_{K_{X,W}}\Big(F_{K_{X}}
(m_{X}),\nonumber \\ &&F_{K_{X}^{'}} (m_{X})\Big), MAC_X
\end{eqnarray}
where
\begin{eqnarray}\label{eq:mac}
MAC_X = MAC\Big{(}K_X,F_{K_{X}} (m_{X})||F_{K_{X}^{'}} (m_{X})\Big{)}
\end{eqnarray}

\begin{algorithm}[t]                    
\caption{SumAgg: $SUM$ Aggregation Algorithm}          
\KwIn{A WSN with set S of $n$ nodes and BS.} \KwOut{$SUM$
aggregation result.}
\begin{algorithmic}
\STATE BS broadcasts $SUM$ aggregation query in the WSN \FOR {$\forall
S_i \in S$}
     \STATE $list_{S_i} = \{ID_{S_i}\}$
     \STATE Sense $m_{S_i}$
     \STATE $DSUM_{S_i} = F_{K_{S_i}}(m_{S_i})$
     \STATE $DSUM^{'}_{S_i} = F_{K^{'}_{S_i}}(m_{S_i})$
      \FOR {$\forall S_j$ that is an immediate child of $S_i$}
        \STATE $DSUM_{S_i} = DSUM_{S_i} + DSUM_{S_j}$
        \STATE $DSUM^{'}_{S_i} = DSUM^{'}_{S_i} + DSUM^{'}_{S_j}$
        \STATE $list_{S_i} = list_{S_i} \cup list_{S_j}$
      \ENDFOR
\ENDFOR \STATE BS sums aggregation of its immediate children nodes. \IF {IPET
check for final aggregation result in the BS passes}
        \STATE return $SUM$
\ELSE
        \STATE Call ComAtt /*Commitment and Attestation Algo.*/
\ENDIF
\end{algorithmic}\label{alg:SumAgg}
\end{algorithm}

As we can see, node $X$ sends its $ID_X$ and an encrypted pair of
its diffused sensed data $m_{X}$ to its parent $W$. $X$ also sends a
pairwise counter $IV_{X,W}$ to protect against replay attacks.
Finally, $X$ sends a MAC of its reading using its private key and
attach it at the end of the packet for authentication purposes as we
shall see later.

The sensor node $W$ receives similar packets from its other
children, i.e. $Y$ and $Z$. Now $W$ needs to aggregate data received
from its children along with its own sensed data $m_{W}$. This is
done through applying the $SUM$ aggregation function as we can see
in the following packet that $W$ sends to its parent $G$:
\begin{eqnarray}
W \longrightarrow G\!\!\!&:&\!\!\! list_{W},IV_{W,G},Enc_{K_{W,G}}
\Big{(}\sum_{S_i \in list_{W}}F_{K_{S_i}}(m_{S_i}), \nonumber \\ &&\sum_{S_i
\in list_{W}}F_{K^{'}_{S_i}}(m_{S_i})\Big{)},MAC_W
\end{eqnarray}
where
\begin{eqnarray}
\!\!MAC_W\!\!\!\!\!&=&\!\!\!\!\! MAC\! \Big(\!K_W,\!\!\!\!\sum_{S_i
\in list_{W}}\!\!\!\!F_{K_{S_i}}(m_{S_i})||\!\!\!\!\!\sum_{S_i \in
list_{W}}\!\!\!\!\!F_{K^{'}_{S_i}}(m_{S_i})\Big) \nonumber \\
&&\oplus MAC_X \oplus MAC_Y \oplus MAC_Z
\end{eqnarray}

Here $list_{W}$ represents the list of all $IDs$ of the children of
$W$ who contributed in the aggregation, including $ID_W$. As we can
see, $W$ sends its $ID_W$ and  $IDs$ of all its children who
contributed in the aggregation, and the aggregated $SUM$ of their
data. As shown above, $W$ sums all pairs of data in order, i.e. all
first elements of every pair are summed together, the same thing
happens to second elements of all pairs. This scenario continues
until the BS receives from every immediate child a packet that
contains the $IDs$ of all nodes participated in the $SUM$
aggregation on the partition rooted by that child, along with its
diffused aggregation pair. The BS then computes the final
aggregation pair $(DSUM, DSUM^{'})$ of diffused summation:
\begin{eqnarray}
(DSUM, DSUM^{'})\!\!\!\!\! &=& \!\!\!\!\!\! \Big(\!\!\sum_{i \in
list_{*}}\!\!\!F_{K_{i}}(m_{S_i}),\!\! \sum_{i \in
list_{*}}\!\!\!F_{K^{'}_{i}}(m_{S_i})\Big{)}
\end{eqnarray}
where \begin{eqnarray}list_* = list_{H} \cup \ldots \cup list_{G}
\cup \ldots \cup list_{Q}\end{eqnarray} The actual values of this
diffused pair $(DSUM, DSUM^{'})$ should refer to the same output,
but since they are diffused differently, they look different.
Because the BS knows $K_{S_i}$ and $K^{'}_{S_i}$ for every node
$S_i$, the BS is able to generate the diffusion values that every
node contributed in the aggregation has used to diffuse its reading,
the BS can revert the pair $(DSUM,DSUM^{'})$ to their actual values.
This is done by finding the summations of all diffusion values that
were applied along the path of aggregation, and using these
summations when applying the reverse diffusion function on counter
parts results $DSUM$ and $DSUM^{'}$:
\begin{eqnarray}
\label{Integ_Eq} (SUM, SUM^{'})\! &=&\!\! \Big{(}DSUM \bar{\odot} \sum_{i
\in list_{*}}D_{i} , \nonumber \\ &&DSUM^{'} \bar{\odot} \sum_{i \in
list_{*}}D^{'}_{i} \Big{)}
\end{eqnarray}

Here, the operand $\bar{\odot}$ refers to the reverse of the
diffusion operation. Now the BS revealed the actual result of $SUM$
and $SUM^{'}$ aggregation, it needs to check the integrity of this
result. The BS checks the equality of reverted pair $SUM$ and
$SUM^{'}$, if they are equal then the aggregation result is accepted
(unless the BS doubts it), otherwise the result is rejected and
attestation process will start to detect the path and the source of
the outliers as explained in Section~\ref{sec:commit}.

The test that uses equation \ref{Integ_Eq} then checks the equality
of resulted pair is called Identical Pair Equality Test (IPET). IPET
is an $O(1)$ heuristic that gives us a quick initial indication about the
integrity of the aggregation result.

\begin{lemma}
The complexity of SumAgg algorithm with data
diffusion is $O(n\ln(n))$ on average, and the BS
needs $O(1)$ to verify the integrity of the final aggregation result.
\end{lemma}
Other aggregation functions like $MEAN$ and $MAX$ can be derived
from above description of $SUM$ aggregation with slight
modifications.

\begin{algorithm}[t]                      
\caption{ComAtt: Commitment and Attestation Algo.}          
\label{alg:ComAtt}                           
\KwIn{$list_*$ (list of $IDs$ of all nodes contributed in an
aggregation), $MAC_{Agg}$ (MAC of final aggregation result)}
\KwOut{$list_L$ (list of $IDs$ of outliers)}
\begin{small}
\begin{algorithmic}                    
\STATE $list_L=\emptyset$, $list_C=\emptyset$ \STATE {$Q = \{S_i:
\forall S_i \in list_* \wedge S_i$ is immediate children of BS\} }
\WHILE {$Q \neq \emptyset$}
 \STATE Pick a node $S_i$ from $Q$
 \STATE $S_i \longrightarrow BS: list_{S_i},IV_{S_i},(DSUM_{S_i},DSUM^{'}_{S_i}),MAC_{S_i}$
 \STATE $MAC^{Calc}_{S_i} =$ Reconstructed $MAC_{S_i}$ in BS using collected data and $MAC_{Agg}$
 \IF    {$MAC^{Calc}_{S_i} \neq MAC_{S_i}$ OR IPET check of $S_i$ packet fails}
        \IF   {$S_i$ is not committed to its previous aggregation packet}
            \STATE $list_C = list_C \cup S_i$
        \ENDIF
        \STATE $list_L = list_L \cup S_i$
        \STATE {$Q = Q \cup \{S_j: \forall S_j \in list_* \wedge S_j$ is immediate children of $S_i$ \} }
 \ENDIF
 \STATE $Q = Q - {S_i}$
\ENDWHILE \FOR {$\forall S_i \in list_L - list_C$} \STATE
$list^{'}_{S_i} = (list_{S_i} - list_L) \cup {S_i}$ \STATE $S_i
\longrightarrow BS:$ \STATE $list^{'}_{S_i},IV_{S_i},$ \STATE
$(\sum_{j \in list^{'}_{S_i}} F_{K_j}(m_j),\sum_{j \in
list^{'}_{S_i}} F_{K^{'}_j}(m_j)),MAC_{S_i}$ \IF    {IPET check of
aggregation pair of $S_i$ passes} \STATE $list_L = list_L - S_i$
\ENDIF

\ENDFOR

RETURN $list_L$
\end{algorithmic}
\end{small}
\end{algorithm}

\section{Commitment and Attestation}\label{sec:commit}

In this section we turn our attention  to verifying sensor's
commitments of aggregation, and attestation for finding outlier or
compromised nodes. Note that we don't consider detecting the case
where a compromised node tries to forge its own data, this is
because such a situation is hard to detect if forged data belongs to
normal data range and this resembles node malfunction. In contrast,
we are interested in detecting compromised nodes that are trying to
forge aggregation data of their non-compromised children. The divide
and conquer algorithm for commitment and attestation ComAtt is presented in
algorithm~\ref{alg:ComAtt}, this algorithm uses IPET check as a heuristic to
reconstruct only those branches of the network MAC tree which are necessary for the attestation process, avoiding  unnecessary reconstruction of the whole MAC tree of the WSN. When the BS discovers that the final
aggregation result fails the IPET check, it starts the attestation
process by adding its immediate children who contributed in the
aggregation to the set Q -which is the set containing nodes to be
tested- for verification. For every node $S_i \in Q$, the BS checks
$S_i$ as follows. The BS asks from every node $S_i \in Q$ to resend
its aggregation packet. The BS then checks the commitment of $S_i$
by constructing its authentication code $MAC^{Calc}_{S_i}$ with the
help of the final aggregation result authentication code $MAC_{Agg}$
and collected data. If $MAC^{Calc}_{S_i}$ is identical to
$MAC_{S_i}$, then the BS knows that $S_i$ is committed to its
previously sent aggregation packet. If $S_i$ is committed and its
aggregation pair passes the IPET check then it is assumed honest
-unless the BS doubts its result as we shall see later- and its
descendants will be excluded from further verifications. On the
other hand, if $S_i$ appeared not to be committed to its previously
sent aggregation, or its aggregation pair fails the IPET test, then
$S_i$ is added to the list of outliers $list_L$, and every children
$S_j$ of $S_i$ is added to the set $Q$ for further investigation.
For the case when commitment test of $S_i$ fails, $S_i$ is also
added to the list of not committed nodes $list_C$.

After processing all nodes in $Q$, $list_L$ will be having suspected
nodes that either not committed or failed the IPET check.
Non-committed nodes in $list_L$ are directly considered dishonest or
compromised without any further investigation. However, it might be
the case that an honest committed node in $list_L$ failed the IPET
check because one or more of its children were compromised. We need
to eliminate such honest nodes from $list_L$, this is done by
further investigation of committed nodes that fail IPET check, i.e.
$S_i \in list_L - list_C$. For every such node $S_i$, the BS
requests a new aggregation of $S_i$ that excludes data from any node
$S_j \in list_L$, that is, the BS is giving $S_i$ a chance to prove
its honesty by finding the aggregation of its only honest children.
If the new aggregation of $S_i$ passes the IPET check, then $S_i$ is
removed from $list_L$, otherwise, it is kept there. Finally, the
ComAtt algorithm returns $list_L$ that contains the set of outliers
or compromised nodes.

\begin{lemma}
The commitment process in ComAtt algorithm
is $O(c \ln n)$ in average for some constant $c$, and $O(n)$ in
the worst case.
\end{lemma}
\begin{proof}
The proof is a direct consequence from the binary tree search algorithm,
considering the height (depth) of aggregation equals $\ln n$ in
average
\end{proof}
\section{Security Analysis}

In this section, we show how our security protocol could be
compared to hop-by-hop and end-to-end protocols in terms of security
level and efficiency of data integrity check.

\subsection{Node Attacks}
We consider the logical hypothesis that a node $S_i$ is attacked by
an intruder (attacker) $I$. This attacker $I$ can gain access to all
information of this node including $K_{S_i}$, $list_{S_i}$ and
$m_{S_i}$. In this case, it can alter the message $m_{S_i}$ to
$m_{I}$ and encrypt it using the key $K_{S_i}$. We show that the
only influence such an attacker can have on final aggregation result
is sending forged aggregation of attacked nodes. If the attacker
attempts to change the aggregation values of its children without
knowing their dual diffusion seeds, then this attempt will be
quickly caught by the IPET test. So, an attacker in this case won't
be able to forge its aggregation except by changing its own reading
$m_{I}$ and aggregations of its children which their dual diffusion
seeds are known to the attacker. That is, if an attacker wants to
forge the aggregation of $n$ nodes and not get caught by IPET, then
this attacker must compromise or acquire private data of $n$ nodes.

\begin{lemma}
Our aggregation protocol represents a security model
against spy-out attacks that is better or at least as good as
hop-by-hop aggregation protocols.
\end{lemma}
\begin{proof}
Our protocol has an advantage over hop-by-hop protocols because of
transient data diffusion. Only when a passive adversary succeeds in
breaking the diffused data of all children of a hop, our protocol
becomes vulnerable to spy-out attacks as any other hop-by-hop
protocol.
\end{proof}

\begin{lemma}
Our protocol performs either more efficient or at least
as good as end-to-end aggregation protocols in checking data integrity.
\end{lemma}

\begin{proof}
In our protocol, we use IPET heuristic to reconstruct the only
necessary branches of the MAC tree for testing data integrity. In
the worst case, we will need to reconstruct the whole MAC tree,
which is the case in end-to-end protocols.
\end{proof}

\section{Conclusions}\label{sec:conclusion}

In this paper, we demonstrated a model for secure data aggregation
in WSNs, which is a blend of hop-by-hop operational efficiency and end-to-end data secrecy.
We showed that this model has low
computational complexity and the BS uses $O(1)$ heuristic to
verify final aggregation result of sensed data and it needs $O(\ln n)$ in average to detect
an attacked node. We plan to perform simulation and further security
analysis of this model in our future work.



\nix{
\scriptsize
\bibliographystyle{plain}
\bibliography{secsensorrefs}

\begin{thebibliography}{17}

\bibitem{Castelluccia05}
C.~Castelluccia, E.~Mykletun, and G.~Tsudik.
\newblock Efficient aggregation of encrypted data in wireless sensor networks.
\newblock In {\em Mobile and Ubiquitous Systems: Networking and Services, 2005.
  MobiQuitous 2005. The Second Annual International Conference on}, pages
  109--117, 2005.

\bibitem{Chan05}
H.~Chan, V.~Gligor, A.~Perrig, and G.~Muralidharan.
\newblock On the distribution and revocation of cryptographic keys in sensor
  networks.
\newblock {\em IEEE Trans. Dependable Secur. Comput.}, 2(3):233--247, 2005.

\bibitem{chan06}
H.~Chan, A.~Perrig, and D.~Song.
\newblock Secure hierarchical in-network aggregation in sensor networks.
\newblock In {\em Proc. 13 ACM conf. on computer and communications security},
  November 2006.

\bibitem{girao05}
J.~Girao, D.~Westhoff, and M.~Schneider.
\newblock {CDA}: Concealed {D}ata {A}ggregation in {W}ireless {S}ensor
  {N}etworks.
\newblock In {\em Proc. 40th International Conference on Communiacations, IEEE
  ICC '05}, Korea, May 2005.

\bibitem{mica2}
M.~Horton, D.~Culler, K.~Pister, J.~Hill, R.~Szewczyk, and A.~Woo.
\newblock {MICA} the commercialization of microsensor motes.
\newblock {\em In Sensors}, 19(4):40--48, April 2002.

\bibitem{hu03}
L.~Hu and D.~Evans.
\newblock Secure aggregation for wireless networks.
\newblock In {\em Symposium on Applications and the Internet Workshops
  (SAINT'03)}, number 384, 2003.

\bibitem{Intanagonwiwat02}
C.~Intanagonwiwat, D.~Estrin, R.~Govindan, and J.~Heidemann.
\newblock Impact of network density on data aggregation in wireless sensor
  networks.
\newblock In {\em Proc. of International Conference on Distributed Computing
  Systems (ICDCS '02)}, Vienna, Austria, July 2002.

\bibitem{madden02}
S.~Madden, M.~Franklinm, J.~Hellerstein, and W.~Hong.
\newblock {TAG}: a tiny aggregation service for ad-hoc sensor networks.
\newblock {\em SIGOPS Oper. Syst. Rev.}, (36(SI):):131--146, May 2002.

\bibitem{Mahimkar2004}
A.~Mahimkar and T.~S. Rappaport.
\newblock Securedav: a secure data aggregation and verification protocol for
  sensor networks.
\newblock In {\em Proc. IEEE Global Telecommunications Conference, 2004.
  GLOBECOM '04.}, volume~4, pages 2175--2179, 2004.

\bibitem{mykletun06}
E.~Mykletun, J.~Girao, and D.~Westhoff.
\newblock Public key based cryptoschemes for data concealment in wireless
  sensor networks.
\newblock In {\em Proc. IEEE International Conference on Communications (ICC
  '06)}, 2006.

\bibitem{PerrigLukKuo2007}
A.~Perrig, M.~Luk, and C.~Kuo.
\newblock Message-in-a-bottle: User-friendly and secure key deployment for
  sensor nodes.
\newblock In {\em Proc. of the ACM Conference on Embedded Networked Sensor
  System (SenSys '07)}, October 2007.

\bibitem{perrig01}
A.~Perrig, R.~Szeczyk, Wen V., D.~Culler, and J.~Tygar.
\newblock {SPINS}: security protocols for sensor networks.
\newblock {\em Mobil Computing and Networkings}, page 189–199, 2001.

\bibitem{przydatek03}
B.~Przydatek, D.~Song, and A.~Perrig.
\newblock {SIA}: Secure information aggregation in sensor networks.
\newblock In {\em Proc. of SenSys 2003}, page 255–265, New York, November 2003.

\bibitem{Raymon08}
D.~Raymond and S.~Midkiff.
\newblock Denial-of-service in wireless sensor networks: Attacks and defenses.
\newblock {\em IEEE Pervasive Computing}, 7(1):74--81, 2008.

\bibitem{Sang06}
Y.~Sang, H.~Shen, Y.~Inoguchi, Y.~Tan, and N.~Xiong.
\newblock Secure data aggregation in wireless sensor networks: A survey.
\newblock In {\em Proc. of the Seventh International Conference on Parallel and
  Distributed Computing, Applications and Technologies (PDCAT '06)}, pages
  315--320, Washington, DC, USA, 2006. IEEE Computer Society.

\bibitem{stojmenovic05}
I.~Stojmenovic.
\newblock Handbook of sensor networks, algorithms and architechtrues.
\newblock {\em Wiley series on parallel and distributed computing}, 2005.

\bibitem{yang06}
Y.~Yang, X.~Wang, and S.~Zhu.
\newblock {SDAP}: a secure hop-by-hop data aggregation protocol for sensor
  networks.
\newblock In {\em Proc. 7th ACM Interational Symposium on Mobile Ad Hoc
  Networking and Computing}, May 2006.

\end{thebibliography}
}

\end{document}